\documentclass[%
 aip,
 amsmath,amssymb,
 reprint,%
]{revtex4-1}

\usepackage{graphicx}
\usepackage{dcolumn}
\usepackage{bm}

\usepackage[utf8]{inputenc}
\usepackage[T1]{fontenc}
\usepackage{mathptmx}
\usepackage{etoolbox}

\makeatletter
\def\@email#1#2{%
 \endgroup
 \patchcmd{\titleblock@produce}
  {\frontmatter@RRAPformat}
  {\frontmatter@RRAPformat{\produce@RRAP{*#1\href{mailto:#2}{#2}}}\frontmatter@RRAPformat}
  {}{}
}%
\makeatother
\begin{document}

\preprint{AIP/123-QED}

\title[]{First implementation of AXUV-based analysis and macro-instability diagnostics on WHAM\\}

\author{K. Shih}
\affiliation{Realta Fusion, Madison, WI 53717, USA}

\author{D. Endrizzi}
\affiliation{Realta Fusion, Madison, WI 53717, USA}

\author{D.A. Sutherland}
\affiliation{Realta Fusion, Madison, WI 53717, USA}

\author{J. Anderson}
\affiliation{Realta Fusion, Madison, WI 53717, USA}
\affiliation{Department of Physics, University of Wisconsin, Madison, WI 53706, USA}

\author{D. Bindl}
\affiliation{Realta Fusion, Madison, WI 53717, USA}

\author{E.L. Claveau}
\affiliation{Realta Fusion, Madison, WI 53717, USA}

\author{C. Everson}
\affiliation{Realta Fusion, Madison, WI 53717, USA}

\author{J. Eickman}
\affiliation{Realta Fusion, Madison, WI 53717, USA}

\author{S. J. Frank}
\affiliation{Realta Fusion, Madison, WI 53717, USA}

\author{E. Marriott}
\affiliation{Realta Fusion, Madison, WI 53717, USA}

\author{E. Penne}
\affiliation{Department of Physics, University of Wisconsin, Madison, WI 53706, USA}

\author{J. Pizzo}
\affiliation{Realta Fusion, Madison, WI 53717, USA}

\author{T. Qian}
\affiliation{Department of Physics, University of Wisconsin, Madison, WI 53706, USA}

\author{J. Viola}
\affiliation{Realta Fusion, Madison, WI 53717, USA}
\affiliation{Department of Nuclear Science and Engineering, MIT, Cambridge, MA 02139, USA}

\author{C. B. Forest}
\affiliation{Realta Fusion, Madison, WI 53717, USA}
\affiliation{Department of Physics, University of Wisconsin, Madison, WI 53706, USA}

\author{D. Yakovlev}
\affiliation{Department of Physics, University of Wisconsin, Madison, WI 53706, USA}

\date{\today}

\begin{abstract}

Absolute extreme ultraviolet (AXUV) diode arrays are widely used in fusion experiments for time-resolved measurements of plasma radiation. We report the first implementation of an AXUV-based analysis framework on the Wisconsin High-Temperature Superconducting (HTS) Axisymmetric Mirror (WHAM). A single, precisely calibrated 20-channel AXUV assembly measures line-integrated plasma emission with $ 100~\mathrm{kHz}$ temporal resolution and $\sim1~\mathrm{cm}$ spatial accuracy across the mid-plane. The data were processed to obtain plasma's statistical moments, yielding time-resolved measurement of the centroid displacement $\Phi(t)$ and effective radius $R(t)$. From the joint covariance of these quantities, we define a macroscopic instability parameter $\chi(t)$, that quantifies large-scale plasma motion and profile evolution directly from AXUV observables. The parameter $\chi$ serves as a compact indicator of global macroscopic instability, decreasing with increasing end-plate bias and exhibiting strong anti-correlation with diamagnetic flux during confinement transitions. These results demonstrate that a single AXUV array can provide quantitative, real-time assessment of macroscopic plasma instabilities, constituting the first demonstration of such capability in a magnetic mirror plasma. Future extensions to multiple arrays will further enhance spatial coverage and enable full-mode tracking in axisymmetric mirror configurations and related fusion devices.

\end{abstract}

\maketitle

\section{\label{sec:level1}Introduction}

The Wisconsin high-temperature superconducting (HTS) Axisymmetric Mirror (WHAM)~\cite{doun_wham} experiment began plasma operations in July 2024 at the University of Wisconsin–Madison Physical Sciences Laboratory (PSL). WHAM is a university–industry collaboration with Realta Fusion and represents a key step towards commercially viable magnetic-mirror fusion. The device comprises a cylindrical vacuum vessel of approximate dimensions $2~\mathrm{m}\times 5~\mathrm{m}$ and employs a pair of HTS mirror magnets capable of producing on-axis fields up to $17~\mathrm{T}$, key parameters shown in Table~\ref{tab:whampp}. At these fields, WHAM targets plasmas with characteristic ion energies of $\sim 10~\mathrm{keV}$. WHAM’s central physics objective is to investigate plasma stability and confinement improvement in the collisionless mirror confinement regime~\cite{doun_wham}, with shear-flow stabilization identified as a hypothesized mechanism for achieving sustained, quiescent confinement in axisymmetric mirror plasmas. This demonstration represents a critical milestone toward reactor-relevant mirror concepts pursued in Realta Fusion’s development pathway~\cite{doun_wham, forest2024prospects, Frank2025}, which aim to realize compact, steady-state fusion systems based on high-beta, axisymmetric mirrors.

Diagnosing macroscopic instability and confinement changes demands time-resolved measurements sensitive to both the plasma's symmetry and macroscopic motion. In this work we focus on absolute extreme ultraviolet (AXUV) diode arrays~\cite{AXUV}, which combine microsecond response, large dynamic range, and compact form factor. These attributes enable dense spatial sampling across the plasma mid-plane and continuous monitoring of plasma emission symmetry, centroid displacement, and profile width. Because AXUV arrays are low-cost and modular, they can be deployed in sufficient channel count to support tomographic analysis while also providing robust, line-integrated observables suitable for real-time interpretation and eventual feedback control.

\begin{table}[b]
\caption{\label{tab:whampp}Key parameters and systems of the WHAM device.}
\begin{ruledtabular}
\begin{tabular}{lc}
Feature & Specification \\
\hline
HTS mirror field   & 17 T (two magnets) \\
Mirror ratio       & $\approx 70$ \\
Plasma radius      & $\sim 0.2\,\mathrm{m}$ \\
ECH system         & 110 GHz, 400–500 kW, 10 ms \\
NBI system         & 25 keV, $\leq 1$ MW \\
Density            & $\sim 3\times 10^{19}\,\mathrm{m^{-3}}$ \\
\end{tabular}
\end{ruledtabular}
\end{table}

\begin{figure}
\includegraphics[width=0.8\linewidth]{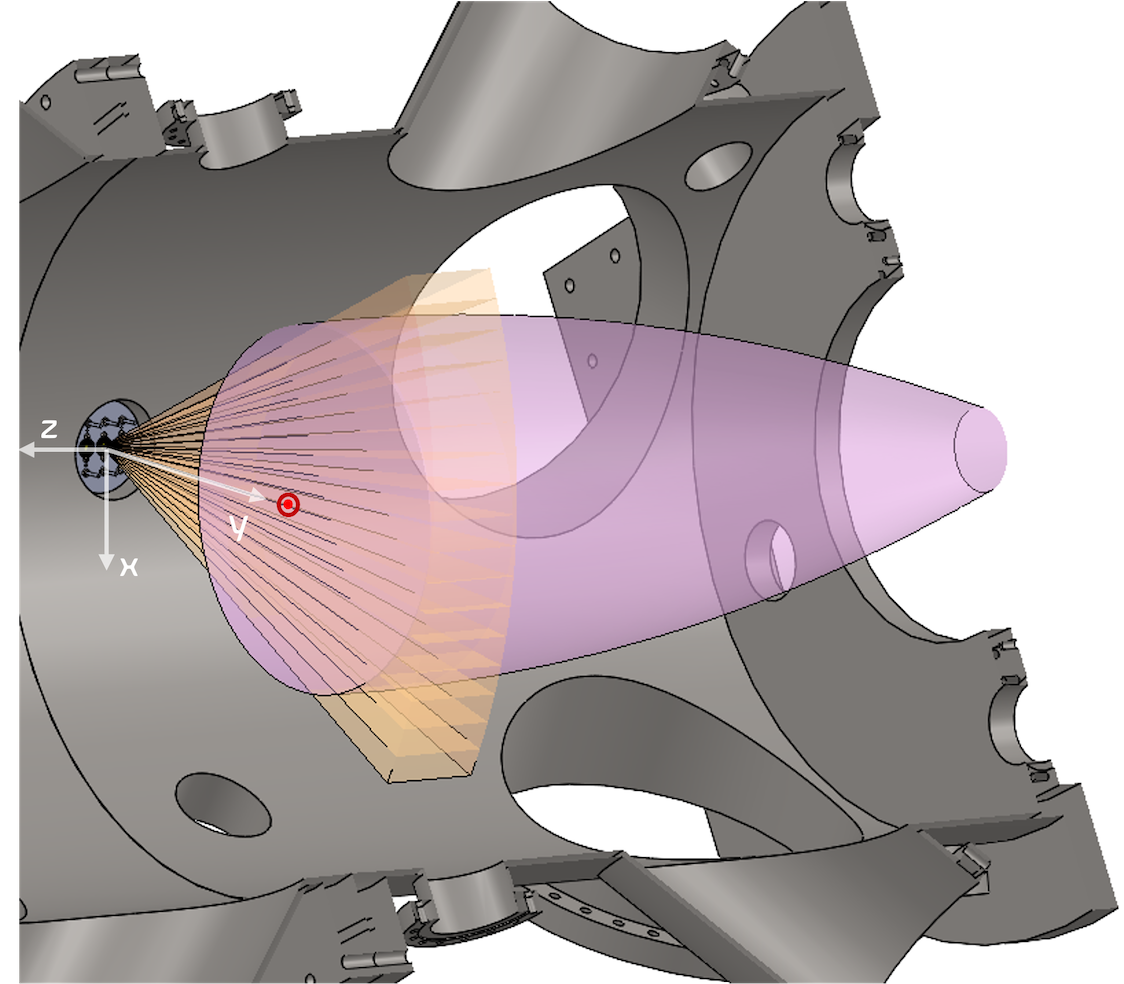}
\caption{\label{fig:AXUV_WHAM} Schematic of the AXUV diagnostic in the WHAM experiment. 
The blue circular plate represents the AXUV assembly; the pink column represents half of the WHAM plasma; 
and the orange fans represent the line-of-sight (LOS) trajectories intersecting the plasma.}
\end{figure}

This paper presents initial AXUV results from the first WHAM campaign. A single 20-channel AXUV assembly was installed at the central-cell mid-plane (Fig.~\ref{fig:AXUV_WHAM}), producing a $20\times 20{,}000$ signal array for a typical $20~\mathrm{ms}$ discharge at $1~\mathrm{MHz}$ sampling. We analyze this data with two complementary objectives. First, we establish a calibrated, geometry-aware processing pipeline that converts raw line-integrated signals into physically referenced profiles across the mid-plane. Second, we extract emission-weighted macroscopic parameters, the plasma centroid $\Phi(t)$ and effective plasma radius $R(t)$. Using their joint statistics, we define an instability parameter $\chi(t)$ that quantifies macroscopic activity directly from AXUV data.

The central contribution is a covariance-based macroscopic instability metric,
constructed solely from single AXUV diode-array observables, that (i) tracks bulk motion and profile broadening in real time, (ii) correlates with diamagnetic-flux signatures of confinement, and (iii) cleanly discriminates inter-discharge trends during bias scans that modify plasma conditions via end-plate biasing. We show that $\chi$ decreases as end-plate bias is raised into regimes associated with improved stability and is strongly anti-correlated with diamagnetic flux during controlled transitions, indicating improved confinement at low $\chi$. This represents the first application of an AXUV-based instability metric in a magnetic mirror device, demonstrating the viability of compact, array-based radiometry for quantitative stability assessment in mirror plasmas.

The paper is organized as follows. Section~\ref{sec:diagnostics} summarizes the AXUV diagnostic design and calibration, establishing geometric accuracy and relative channel sensitivity. Section~\ref{sec:analysis} details the data-analysis methods, including symmetry handling and the moment-based definitions of $\Phi(t)$ and $R(t)$ that support the instability analysis. Section~\ref{sec:stability} introduces the macroscopic instability parameter $\chi$, developing its physical interpretation, and presents representative experimental results, including both multi-discharge bias scans and intra-discharge comparisons with diamagnetic flux, from  flux-loop measurements~\cite{ pustovitov2020diagnostic}. Conclusions and outlook are given at the end.

\section{\label{sec:diagnostics}Diagnostic Overview}

The AXUV diagnostic provides the experimental foundation for all subsequent analyses in this paper. It functions as a $20 \times 1$–resolution camera: twenty spatial samples across the plasma cross section, each integrating the emission profile along the mirror axis. The resulting array of signals captures the emissivity distribution across the WHAM mid-plane with high temporal resolution, enabling quantitative studies of plasma symmetry, displacement, and profile evolution.

Reliable interpretation of this data requires accurate geometric and sensitivity calibration. Each diode’s position and relative response must be known with respect to the WHAM magnetic axis to convert raw voltage traces into physically meaningful emissivity projections. The AXUV diodes used in WHAM are photosensitive across a broad photon-energy range ($\sim1\,\mathrm{eV}$–$1\,\mathrm{keV}$). To suppress unwanted optical reflections and isolate soft-x-ray emission, a $250\,\mathrm{nm}$ aluminum filter was installed in front of the diodes during the initial experimental campaign.

\subsection{\label{sec:axuv_design}AXUV Assembly Design}

\begin{figure}[h!]
\centering
\includegraphics[width=0.8\linewidth]{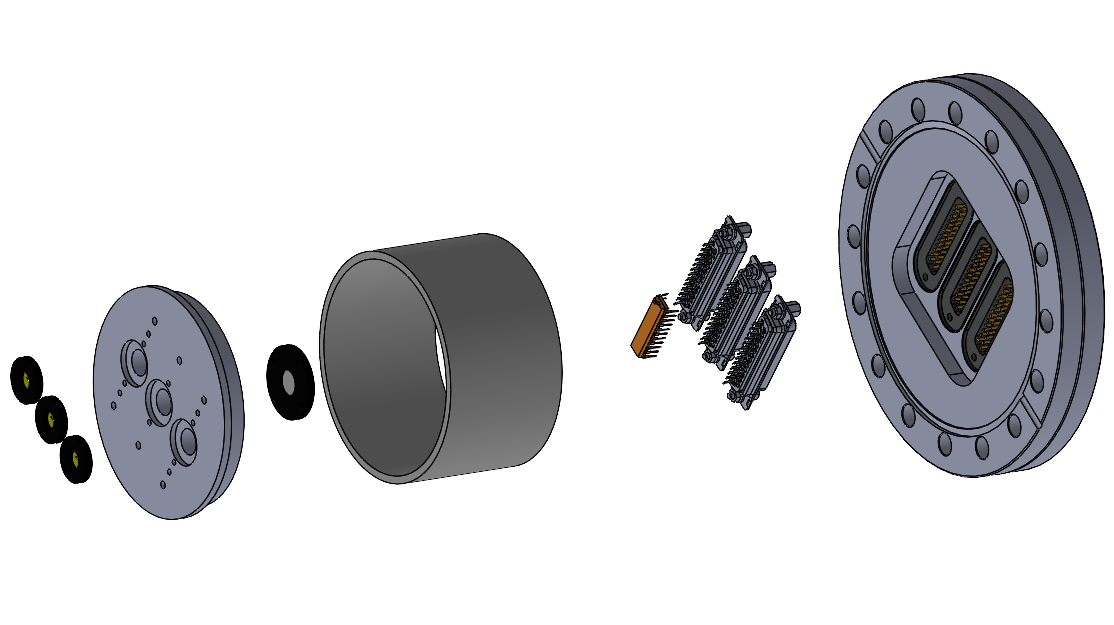}
\caption{\label{fig:AXUV_Assembly} Exploded view of the AXUV assembly. Plasma emission is imaged by the precision gold slit (Layer~1, from the left), passes through the aluminum filter (Layer~3), and is projected onto the AXUV photodiode array (Layer~5). (An intermediate mechanical spacer/support forms Layer~2, ~4 and sets the slit–detector distance.)}
\end{figure}

The AXUV assembly operates as a slit-pinhole camera. Photons emitted from the plasma pass through a precision gold slit and form a line image on the diode array. A schematic of the diagnostic assembly is shown in Fig.~\ref{fig:AXUV_Assembly}. By inserting interchangeable filters between the slit and the diode array, emissivity can be measured across multiple photon-energy bands. Although the design accommodates three simultaneous spectral channels, only the central diode array, equipped with the $250\,\mathrm{nm}$ aluminum filter, is used in the present campaign.

The solid angle viewed by each diode is determined by the slit geometry and the relative slit-to-array alignment. Assuming uniform emissivity along the mirror axis, the measured diode signal can be expressed as in Eq.~(\ref{eq:axuvI}), where $I_i(t)$ is the measured signal and $C_{i0}$ is the relative calibration factor of diode $i$ with respect to the central diode ($i=0$). Here, $S(r,\theta,t)$ denotes the emissivity distribution, and $\bar{S_i}(r,t)=\langle S(r,\theta_i,t)\rangle_{\Delta_i}$ represents the emissivity averaged over the acceptance cone $\Delta_i$ centered at $\theta_i$.

\begin{align}\label{eq:axuvI}
I_i(t) &= C_{i0} \int \bar{S_i}(r,t)\,dr, \qquad i\in\mathbb{Z}
\end{align}

Because the $1/R^2$ falloff of radiated power and the $R^2$ increase of solid angle cancel exactly in the local geometry, each diode effectively records a line integral of emissivity across the plasma radius. Residual geometric differences among channels are absorbed into the calibration coefficient $C_{i0}$, determined experimentally as described below.

\subsection{\label{sec:axuv_calibration}Diagnostic Calibration}

\begin{figure}[h!]
\centering
\includegraphics[width=0.8\linewidth]{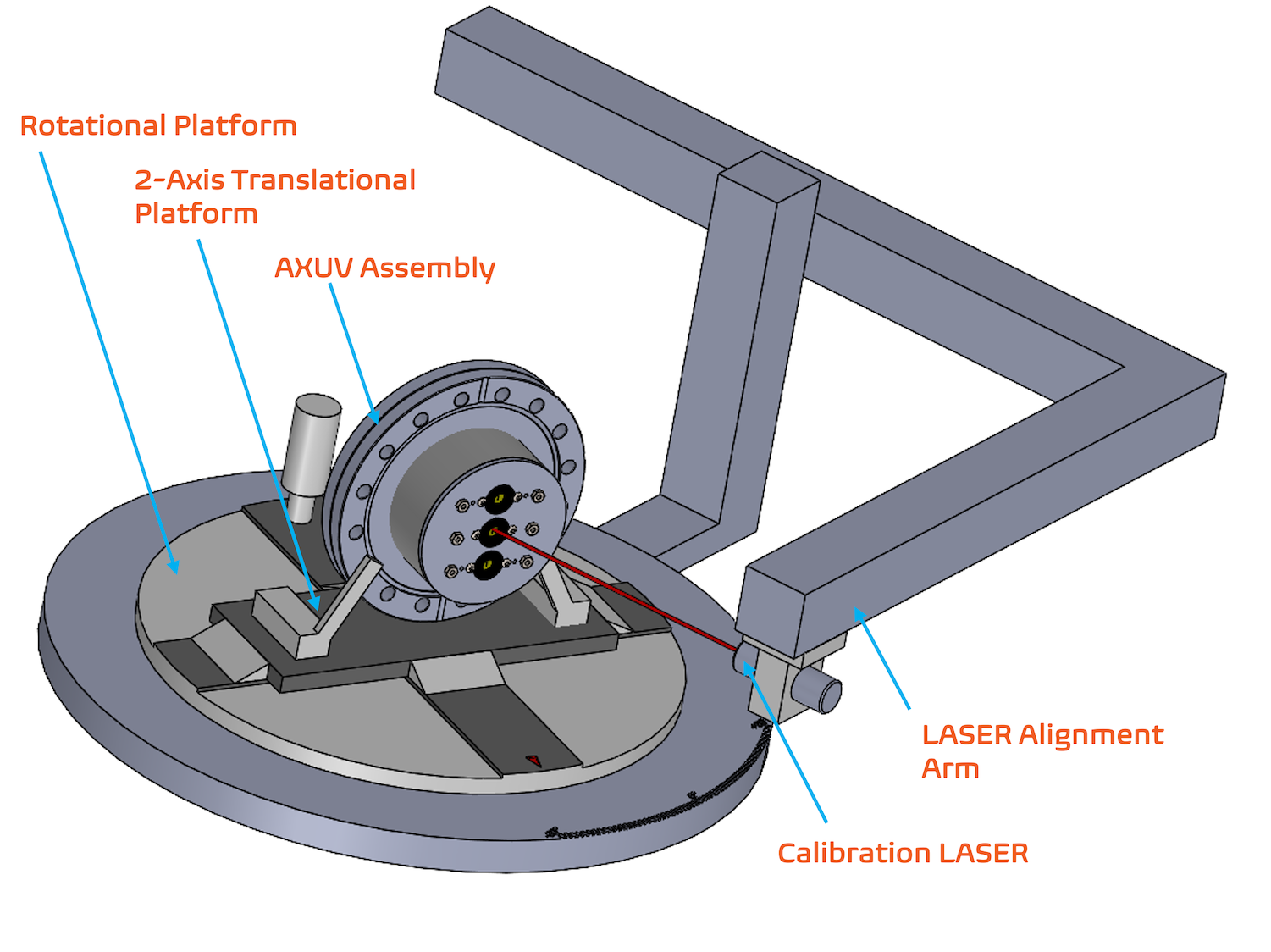}
\caption{\label{fig:AXUV_calibration} Calibration platform for the AXUV diagnostic. The AXUV assembly is mounted on a rotation and two-axis translation stage, while a fixed laser alignment arm holds the light source.  
By adjusting the platform orientation and position, the focused laser beam can be directed onto the slit from different lines-of-sight, enabling precise calibration of each diode’s viewing geometry and relative response.}
\end{figure}

The relative efficiency factor $C_{i0}$ accounts for variations in solid angle, electronic impedance, cable transmission, and diode quantum efficiency. Assuming uniform spectral response across the array, these systematic effects are determined via a relative calibration using a laser light source as a reference.  A schematic of the calibration apparatus is shown in Fig.~\ref{fig:AXUV_calibration}.

During calibration, the AXUV assembly is fixed at the center of a rotatable platform, while the laser source is mounted on an adjustable arm that rotates around it.
The platform allows both rotation and fine two-dimensional translation so that the focused laser beam can be aligned precisely through the slit along different lines of sight (LOS).
Alignment is achieved by adjusting the platform position so that the slit lies at the laser focal point and at the geometric center of rotation. Proper focus is verified by monitoring the laser spot on the slit as the arm rotates: motion of the spot in phase with rotation indicates that the slit lies upstream of the focus, out-of-phase motion indicates it is downstream, and a stationary spot confirms correct alignment at the focal plane.

\begin{figure}[h!]
\centering
\includegraphics[width=0.8\linewidth]{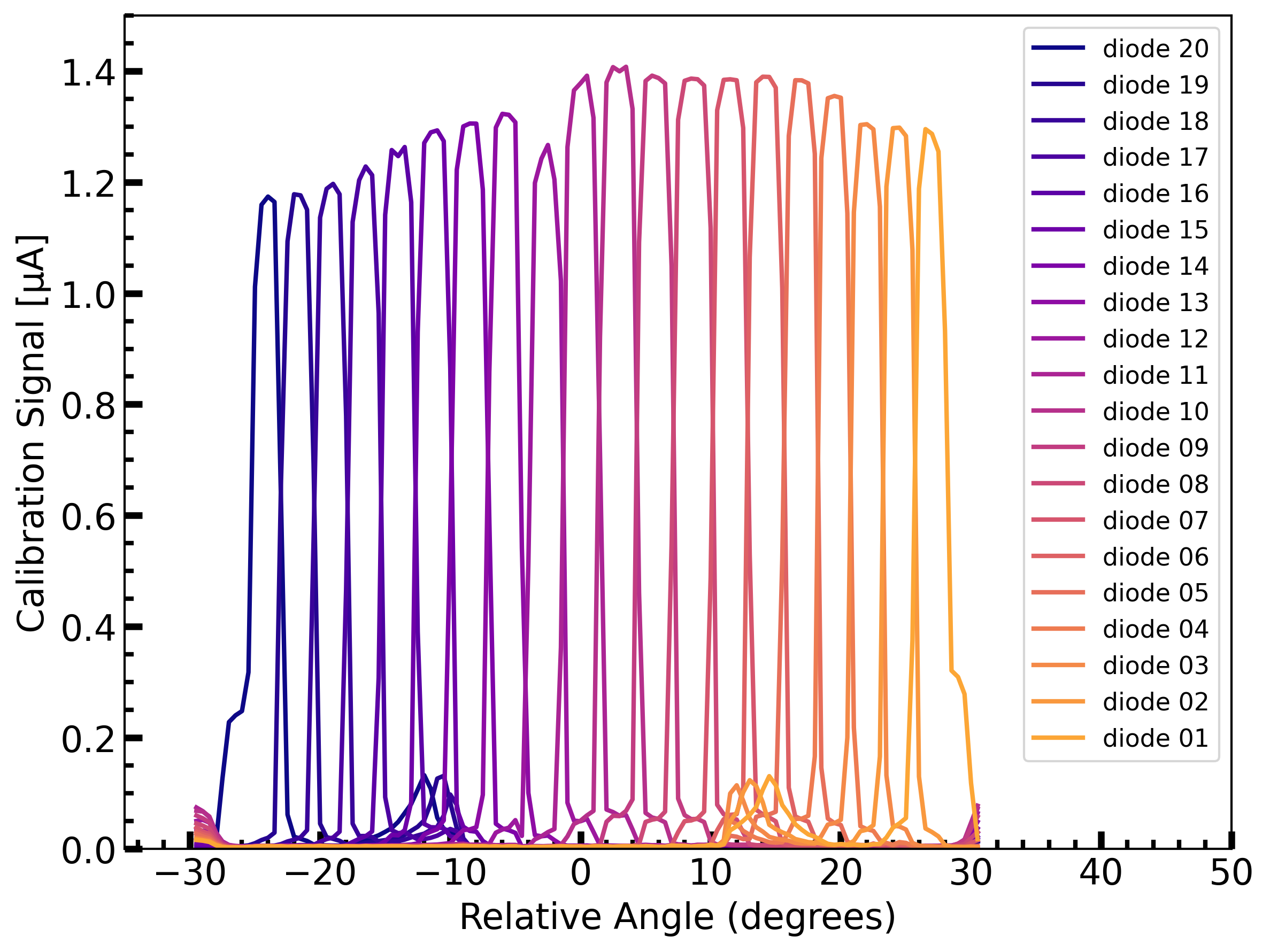}
\caption{\label{fig:cal_curve} Calibration curves for all 20 AXUV diodes, showing the relative response as a function of viewing angle. Each curve corresponds to an individual diode, with colors representing the diode index, where $i = 11 - \text{diode index}$}
\end{figure}

An example calibration curve is shown in Fig.~\ref{fig:cal_curve}. From the relative peak intensities, the product $C_{i0}R_i$ is obtained, where $R_i$ is the channel impedance. The relative angular position $\theta_{i0}=\theta_i-\theta_0$ of each diode is extracted from the signal geometry. After calibration, all channels are referenced to the central diode, aligned to the WHAM vessel center and nominal magnetic axis. The absolute angular position $\phi_0$ is re-established at each installation using the same optical alignment procedure, while the remaining diode positions follow from $\phi_i=\theta_{i0}+\phi_0$.

This calibration ensures that subsequent analysis, detailed in Sec.~\ref{sec:analysis}, uses accurately referenced, sensitivity-corrected data. The resulting set of time-resolved, line-integrated emissivity measurements forms the quantitative basis for the macroscopic instability analysis presented in Sec.~\ref{sec:stability}.

\section{\label{sec:analysis}Data Analysis Methods} 

This section outlines the data-processing framework used to extract physically meaningful plasma parameters from the AXUV diagnostic signals.  
The objective is to translate the raw, line-integrated diode array data into quantities that capture the macroscopic dynamics of the WHAM plasma, ultimately forming the foundation for the instability analysis presented in Sec.~\ref{sec:stability}.  
The procedure combines geometric calibration of the viewing system with statistical moment analysis of the time-resolved emissivity profiles.

\subsection{\label{sec:tomography}Diagnostic Geometry and Signal Formation}

The calibrated AXUV assembly provides line-integrated soft–x-ray emissivity signals $I_i(t)$ along a fan of chords at viewing angles $\phi_i$ intersecting the WHAM plasma column.  
A local coordinate system is defined with the AXUV slit at the origin, the $z$-axis aligned with the mirror axis (pointing north), and the $(x,y)$ plane coincident with the plasma mid-plane.  
The $y$-axis points toward the WHAM vessel center, coinciding with the magnetic axis (Fig.~\ref{fig:AXUV_WHAM}).  
To maintain a right-handed convention, negative viewing angles $\phi$ are defined in the $+x$ quadrant.  
Each diode channel therefore corresponds to a known line of sight characterized by its impact parameter
\begin{equation}
\label{eq:b}
b_i = -D\sin(\phi_i)
\end{equation}
where $D$ is the distance from the magnetic axis to the AXUV slit.  
The sign convention ensures that $b$ increases with $+x$ and recovers the correct orientation in the limit of parallel projection ($D\to\infty$).

Time-resolved fan-beam signals are mapped to their corresponding impact parameters using Eq.~(\ref{eq:b}). The raw data set ${b_i, I_i}$ normally exhibits finite edge values. To regularize the analysis near the plasma boundary, the measured profile on each side is extended with independent Gaussian decays that smoothly approach zero at the outer edge. Those edges corresponding to the limiter plus the finite–Larmor–radius (FLR) position, $R_L = 25~\mathrm{cm}$. The left ($-$) and right ($+$) extensions are defined in Eq.~(\ref{eq:Gfalling_pm}), where $A_{\pm}$ and $\mu_{\pm}$ are chosen to match both the value and local slope at the outermost measured points $b_{\pm}$.  
The Gaussian widths are set by the distance to the limiter, as shown in Eq.~(\ref{eq:sigma_pm}), with $N_{\sigma}=2$ used in the present analysis. Consequently, resulting the composite profile is defined on $b_j \in [-R_L, R_L]$ as in Eq.~(\ref{eq:If}).

\begin{equation}
\label{eq:Gfalling_pm}
I_{\mathrm{ext}}^{(\pm)}(b)
 = A_{\pm}\exp\!\left[-\frac{(b-\mu_{\pm})^2}{2\sigma_{\pm}^2}\right]
\end{equation}

\begin{equation}
\label{eq:sigma_pm}
\sigma_{\pm} = \frac{|R_L \mp b_{\pm}|}{N_{\sigma}}
\end{equation}

\begin{equation}
\label{eq:If}
I_f(b_j) =
\begin{cases}
I_{\mathrm{ext}}^{(-)}(b_j), & -R_L \le b_j < b_{-} \\[4pt]
I(b_{j=i}), & b_{-} \le b_{j=i} \le b_{+} \\[4pt]
I_{\mathrm{ext}}^{(+)}(b_j), & b_{+} \le b_j \le R_L \\[4pt]
\end{cases}
\end{equation}

Because the native chord spacing is nonuniform in $b_i$, and the Gaussian extensions introduce additional edge points, the composite profile $I_f(b_j)$ is interpolated onto a uniform impact-parameter grid $\{b\}$ prior to further analysis.

In summary, the raw detector geometry $\{b_i, I_i(b_i)\}$ is first expanded into the edge-extended geometry $\{b_j, I_f(b_j)\}$ and subsequently remapped onto the uniform grid $\{b, I_f(b)\}$, which fully spans the plasma cross section and provides a consistent basis for statistical moment analysis.

\subsection{\label{sec:macro_params}Macroscopic Plasma Parameter Analysis}
Once the composite profile $I_f(b)$ is defined on a uniform impact-parameter grid, the data can be directly analyzed to extract global plasma parameters that characterize the macroscopic evolution of each discharge. From the time-resolved diode array measurements $\{b, I_f(b,t)\}$, two statistical quantities are evaluated: the plasma centroid position $\Phi(t)$ and the effective plasma radius $R(t)$. The centroid $\Phi(t)$ represents the emission-weighted displacement of the plasma column and is defined as the first moment of the measured emissivity signal, shown in Eq.~(\ref{eq:centroid}).
\begin{equation}
\label{eq:centroid}
\Phi(t) = \frac{\sum_b I_f(b,t) \,b}{\sum_b I_f(b,t)}
\end{equation}

This quantity reflects the instantaneous asymmetry of the emissivity distribution with respect to the magnetic axis and therefore enables time-resolved tracking of the plasma bulk location relative to the AXUV assembly.  
To further interpret this motion, the measured signal is decomposed into symmetric and anti-symmetric components, $I_{\mathrm{sym}}$ and $I_{\mathrm{anti}}$, respectively, defined with respect to the magnetic center projection (e.g., $I_{\mathrm{sym}}(b,t) = [I_f(b,t) + I_f(-b,t)]/2$).  
The anti-symmetric component $I_{\mathrm{anti}}(b,t)$ captures odd-parity features of the emissivity profile, typically associated with bulk $m\!\approx\!1$ motion of the plasma column.  
This relationship can be expressed as
\begin{equation}
\label{eq:m1}
\Phi(t) = \frac{2N_a(t)}{N(t)}\,\Phi_a(t)
\end{equation}
where $N_a$ and $N$ denote the zeroth moments of $I_{\mathrm{anti}}(b\geq0,t)$ and $I_f(b,t)$, respectively, and $\Phi_a$ represents the mean location of $I_{\mathrm{anti}}(b\geq0,t)$, which is computed directly from the measured data.  
The ratio $2N_a/N$ quantifies the relative amplitude of the $m=1$ population, while $\Phi/\Phi_a$ provides a direct measurement of the normalized $m=1$ mode strength in emissivity space.

Similarly, the plasma radius $R(t)$, defined relative to the plasma center, characterizes the effective width of the emissivity distribution and is given by the second moment of the total signal $I_f(b,t)$, as shown in Eq.~(\ref{eq:radius}). The shifting term $\Phi^2(t)$ arises from referencing the moment to the plasma center; this term should be omitted when defining $R(t)$ relative to the magnetic center. In that case, $R(t)$ depends solely on the symmetric component $I_{\mathrm{sym}}(b,t)$ . Additionally, the scaling factor of 4 in Eq.~(\ref{eq:radius}) corresponds to the exact normalization for a uniform disk distribution. One may alternatively choose a different normalization depending on convention or specific application, but in this work we retain this definition for consistency.

\begin{equation}
\label{eq:radius}
R^2(t)=4\frac{\sum_bI_f(b,t)\,(b-\Phi)^2}{\sum_b I_f(b,t)}
\end{equation}

\subsection{Summary and Connection to Instability Analysis}

The time-resolved quantities $\Phi(t)$ and $R(t)$ provide compact, emission-weighted measurements of plasma displacement and profile broadening, respectively; an example is shown in Appendix Fig.~\ref{fig:axuv_overview}. Fluctuations in these parameters reflect the collective motion of the emitting plasma column and provide a natural basis for defining a macroscopic instability metric. In the next section, we introduce the covariance-based instability parameter $\chi$, constructed from the joint dynamics of $\Phi(t)$ and $R(t)$, to quantify global plasma instability and its dependence on confinement conditions.

\section{\label{sec:stability}Macroscopic Instability Parameter $\chi$ and Experimental Results}

To characterize the overall macroscopic instability, we define a state parameter $\chi(t)$ that quantifies the areal spread of joint fluctuations in the plasma centroid and effective radius. Over a chosen analysis window $\Delta$, the temporal covariance matrix of $(\Phi,R)$ is
\begin{equation}
\Sigma \equiv
\begin{pmatrix}
\sigma_{\Phi\Phi} & \sigma_{\Phi R} \\
\sigma_{R\Phi} & \sigma_{RR}
\end{pmatrix},
\qquad
\sigma_{ab} \equiv \langle \delta a\,\delta b\rangle_\Delta
\end{equation}
where $\delta a \equiv a-\langle a\rangle_\Delta$ and $\langle\cdot\rangle_\Delta$ denotes a time average over the window $\Delta$. We then define
\begin{equation}
\label{eq:chi}
\chi^2 \equiv \det\Sigma
\end{equation}
Geometrically, $\chi$ measures the area of the covariance ellipse in the $(\Phi,R)$ plane: small, stationary $\chi$ indicates quiescent macroscopic behavior, whereas a growing $\chi$ signals increasing centroid and radius fluctuations consistent with MHD-like motion. This construction offers a compact, AXUV-only proxy for tracking bulk motion and profile broadening in real time. A representative time trace of $\Phi(t)$, $R(t)$, and $\chi(t)$ for a typical WHAM discharge is shown in the left bottom plot of  Fig.~\ref{fig:axuv_overview}.

We assume the dominant centroid and radius responses can be represented as in Eq.~(\ref{eq:cR}), where $\gamma_\Phi$ and $\gamma_R$ are the respective growth (or damping) rates, and $G_\Phi(t)$ and $G_R(t)$ are bounded oscillatory functions. Consequently, the entries of $\Sigma$ scale as $e^{2(\gamma_\Phi+\gamma_R)t}$, up to oscillatory factors from $G_\Phi$ and $G_R$. Furthermore, the instability parameter $\chi(t)$ can be expressed as in Eq.~(\ref{eq:chi_growth}), where $\chi_0 \sim \Phi_0 R_0$ and $G(t)$ is a bounded oscillatory function of order unity.

\begin{equation}
\label{eq:cR}
\Phi(t) = \Phi_0\,e^{\gamma_\Phi t}\,G_\Phi(t), \qquad
R(t) = R_0\,e^{\gamma_R t}\,G_R(t)
\end{equation}

\begin{equation}
\label{eq:chi_growth}
\chi(t) = \chi_0\,e^{(\gamma_\Phi+\gamma_R)t}\,G(t)
\end{equation}

Thus, $\chi$ grows when the net growth rate $\gamma_\Phi+\gamma_R>0$ and decays when $\gamma_\Phi+\gamma_R<0$, providing a direct physical interpretation of the macroscopic instability dynamics captured by the AXUV observables. In this framework, decreasing $\chi$ with increasing end-plate bias corresponds to reduced global mode growth and improved macroscopic stability.

\begin{figure}[t]
  \centering
  \includegraphics[width=0.9\linewidth]{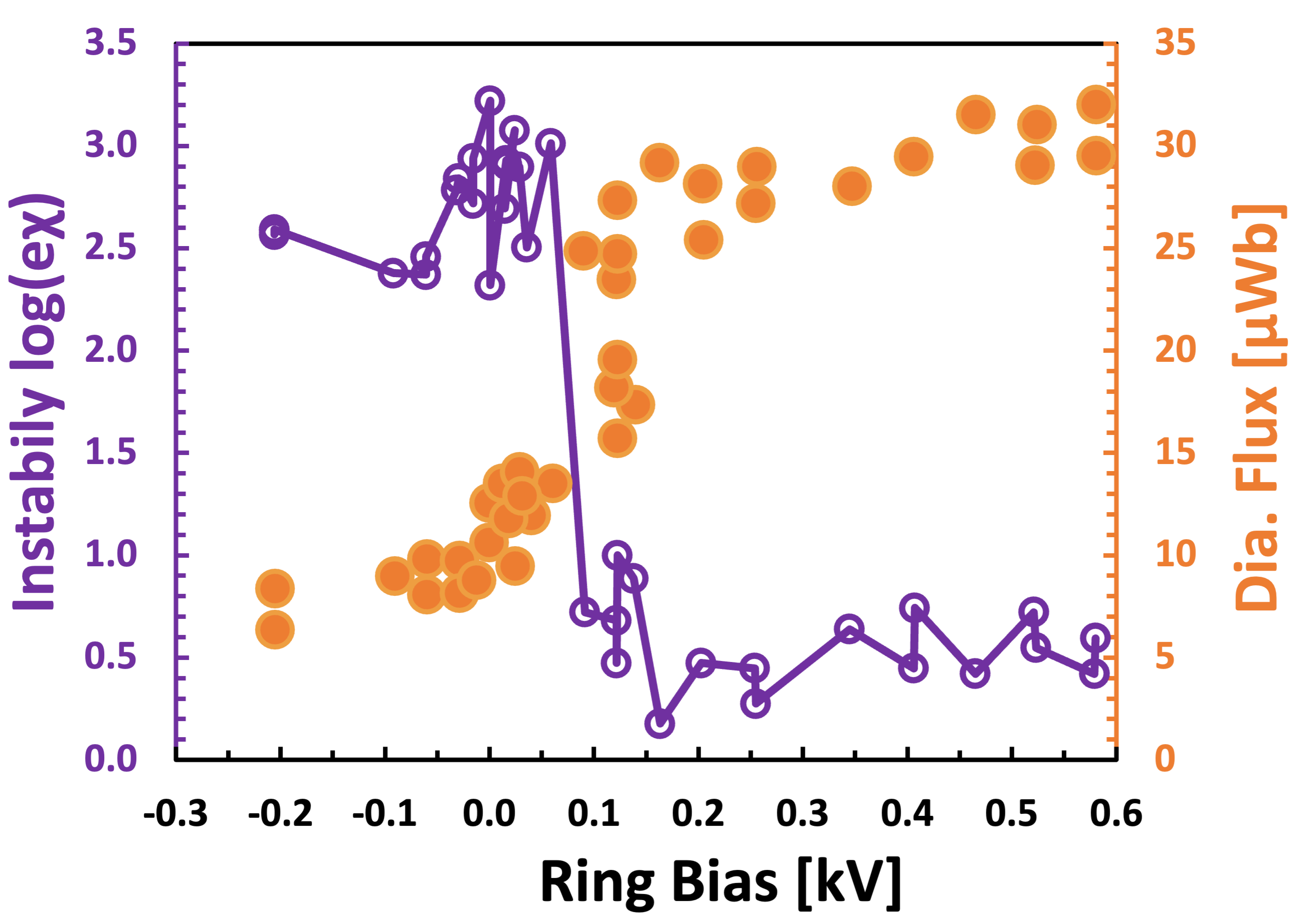}
  \caption{\label{fig:bias_scan}   The inter-discharge bias scan, representative $\log(e\,\chi)$ values reconstructed from AXUV data for different end-plate bias settings, shown together with flux-loop stored-energy proxies. The factor $1/e\,\mathrm{mm^2}$ provides a dimensional normalization such that $\log(e\,\chi)$ is dimensionless. 
  Increasing end-plate bias correlates with reduced $\log(e\,\chi)$ and enhanced stored magnetic energy, indicating improved macroscopic stability and confinement.}
\end{figure}

To explore this connection experimentally, a series of WHAM discharges was executed in which the end-plate bias was systematically varied, modifying plasma boundary conditions and confinement properties~\cite{Qian2025}. Figure~\ref{fig:bias_scan} compares individual discharge reconstructions of $\chi$ for representative bias settings. As the applied bias is increased, $\chi$ transitions from high to low values, delineating bias regimes associated with reduced macroscopic activity and improved confinement. The trend is consistent with simultaneous flux-loop measurements of stored magnetic energy shown alongside in Fig.~\ref{fig:bias_scan}. Although bias-driven shear flow is a leading candidate mechanism~\cite{beklemishev2010vortex} for this behavior, the present analysis focuses on macroscopic instability signatures inferred from AXUV observables.

\begin{figure}[t]
  \centering
  \includegraphics[width=1\linewidth]{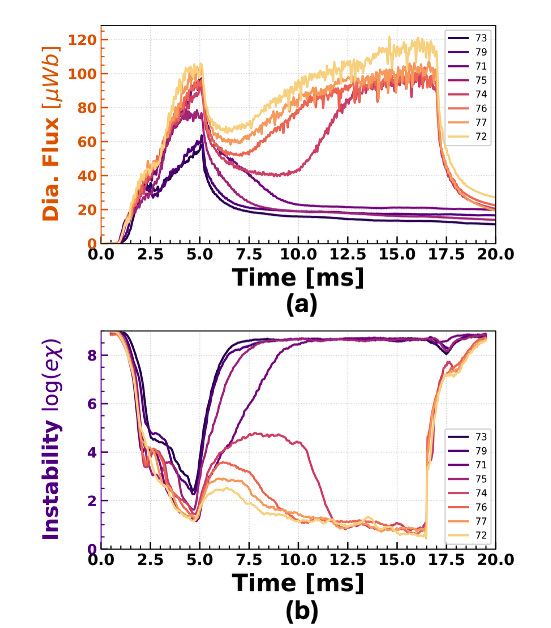}
  \caption{\label{fig:chi_flux} Intra-discharge comparison between $\chi(t)$ and diamagnetic flux measurements. (a) Diamagnetic flux signals obtained from flux-loop diagnostics showing confinement transitions induced by neutral-pressure variation. (b) Corresponding $\chi(t)$ traces derived from AXUV data, plotted as $\log(e\,\chi)$. Different WHAM discharges are shown as distinct colored traces, with shot numbers labeled in the legend (e.g., 250324000-series). The confinement transition occurs between discharges 74 and 75, corresponding to a change in neutral-pressure setting. Periods of low $\chi$ coincide with increased diamagnetic flux (improved confinement), whereas high $\chi$ corresponds to enhanced macroscopic activity and reduced confinement.}
\end{figure}

Complementing the inter-discharge scan, an intra-discharge study compared $\chi(t)$ with diamagnetic flux measurements of stored magnetic energy under controlled transitions. Keeping the bias configuration fixed, the neutral gas pressure was varied to induce transitional bias-current conduction. Under these conditions, the plasma bifurcates into distinct confinement states sustained by NBI injection, as shown in Fig.~\ref{fig:chi_flux}(a). The corresponding $\chi(t)$ traces in Fig.~\ref{fig:chi_flux}(b) closely track the diamagnetic flux: low $\chi$ coincides with higher stored energy (improved confinement), while elevated $\chi$ accompanies enhanced macroscopic activity and a reduction in energy. This strong anti-correlation demonstrates that $\chi$ serves as a convenient instability gauge derivable directly from AXUV measurements.

For comparison, one can form $\Pi \equiv \prod_m \sqrt{\sigma_{mm}}$,
which scales as $\Pi \propto e^{\sum_m \gamma_m t}$ when multiple modes grow independently. However, $\Pi$ does not vanish for perfectly correlated modes and, therefore, cannot distinguish truly independent instabilities from harmonics or sidebands of a single structure. In contrast, the determinant-based measure $\chi=\sqrt{\det \Sigma_{mn}}$ naturally excludes such correlations, yielding a more robust quantification of the independent macroscopic fluctuation content, typically in this single AXUV detector setting.

In summary, $\chi$ provides a physically interpretable, experimentally convenient metric for monitoring equilibrium quality and for assessing the impact of end-plate biasing on macroscopic stability in WHAM, with consistent signatures observed across both inter-discharge bias scans and intra-discharge confinement transitions.

\section{\label{sec:conclusion}Conclusion}

A fast, single-array AXUV diagnostic has been developed and applied to characterize macroscopic plasma behavior in the WHAM experiment.
The system combines precise geometric calibration with an emission-weighted analysis pipeline that converts time-resolved diode signals into physically referenced statistical moments of the plasma emissivity, specifically the centroid and effective radius.

From these moments, the macroscopic instability parameter $\chi(t)$ was defined and evaluated for each discharge, providing a compact, dimensionless measure of large-scale plasma motion and profile evolution.
The instability parameter $\chi$ serves as a quantitative link between radiation dynamics and global confinement. Comparison with independent flux-loop measurements demonstrates a strong anti-correlation: low $\chi$ corresponds to enhanced confinement and high stored energy, while elevated $\chi$ reflects macroscopic activity and reduced stability, Fig.~\ref{fig:chi_flux}.

Across bias-scan and intra-discharge studies, $\chi$ consistently decreases under conditions associated with improved macroscopic stability, demonstrating its utility as a real-time indicator of bias-driven stabilization in WHAM. The results establish that even a single AXUV array can deliver real-time, discharge-resolved assessments of plasma equilibrium and macroscopic instability. Because the method relies only on calibrated diode signals and moment-based analysis, it operates with minimal computational overhead and is well suited for integration into fast real-time feedback control loops.

Looking ahead, the diagnostic platform has been expanded to include additional AXUV arrays and multiple photon-energy bands, introducing an energy-resolved dimension to the emissivity analysis. These upgrades will enable spatial and spectral separation of core, edge, and impurity radiation, improving the physical interpretability of $\chi$ and related quantities. Together, these developments position the AXUV diagnostic suite as a cornerstone of WHAM’s stability and confinement studies, and as a scalable framework for future axisymmetric mirror and fusion devices.


\begin{acknowledgments}
This work was funded in part by Realta Fusion and the Advanced Research Projects Agency-Energy (ARPA-E), U.S. Department of Energy under Award Numbers DE-AR0001258, DE-AR0001261, and U.S. Department of Energy under Award Number DE-FG02-ER54744.
\end{acknowledgments}

\appendix
\section{\label{app:abel}Abel inversion and emissivity reconstruction}

For completeness, this appendix summarizes the Abel inversion~\cite{Buie1996} used to reconstruct the radial emissivity from the symmetrized AXUV data. Considering only the symmetric component $I_{\mathrm{sym}}(b,t)$, each chord is modeled as a parallel, line–integrated projection of the emissivity $\bar{S}(\rho,t)$ under an assumption of cylindrical symmetry. Here $\rho$ is the radial coordinate measured from the WHAM magnetic axis, and its correspondence with the impact parameter $b$ in the AXUV frame follows from the viewing geometry in Eq.~(\ref{eq:b}). Under these assumptions, the projection takes the standard Abel form
\begin{equation}
\label{eq:Abel}
I_{\mathrm{sym}}(b,t)
= 2\!\int_b^{R_L} \frac{\rho\,\bar{S}(\rho,t)}{\sqrt{\rho^{2}-b^{2}}}\,d\rho\, .
\end{equation}

Prior to inversion, the measured profiles are symmetrized and smoothly continued to the outer radius $R_L$ using a Gaussian edge extension. We then invert Eq.~(\ref{eq:Abel}) with the regularized basex algorithm~\cite{Dribinski2002} to obtain $\bar{S}(\rho,t)$.  
This furnishes an axisymmetric realization of the general measurement relation in Eq.~(\ref{eq:axuvI}), mapping the line–integrated diode signals $I_i(t)$ onto a local emissivity distribution in the WHAM coordinate frame.  
Compared with discrete Tikhonov–type matrix inversions, the Abel approach exploits the cylindrical symmetry explicitly and requires fewer a priori smoothness assumptions; the Gaussian continuation plays a role analogous to a stabilizer while preserving the analytic kernel structure of the Abel transform.

\begin{figure}[t]
  \centering
\includegraphics[width=1\linewidth]{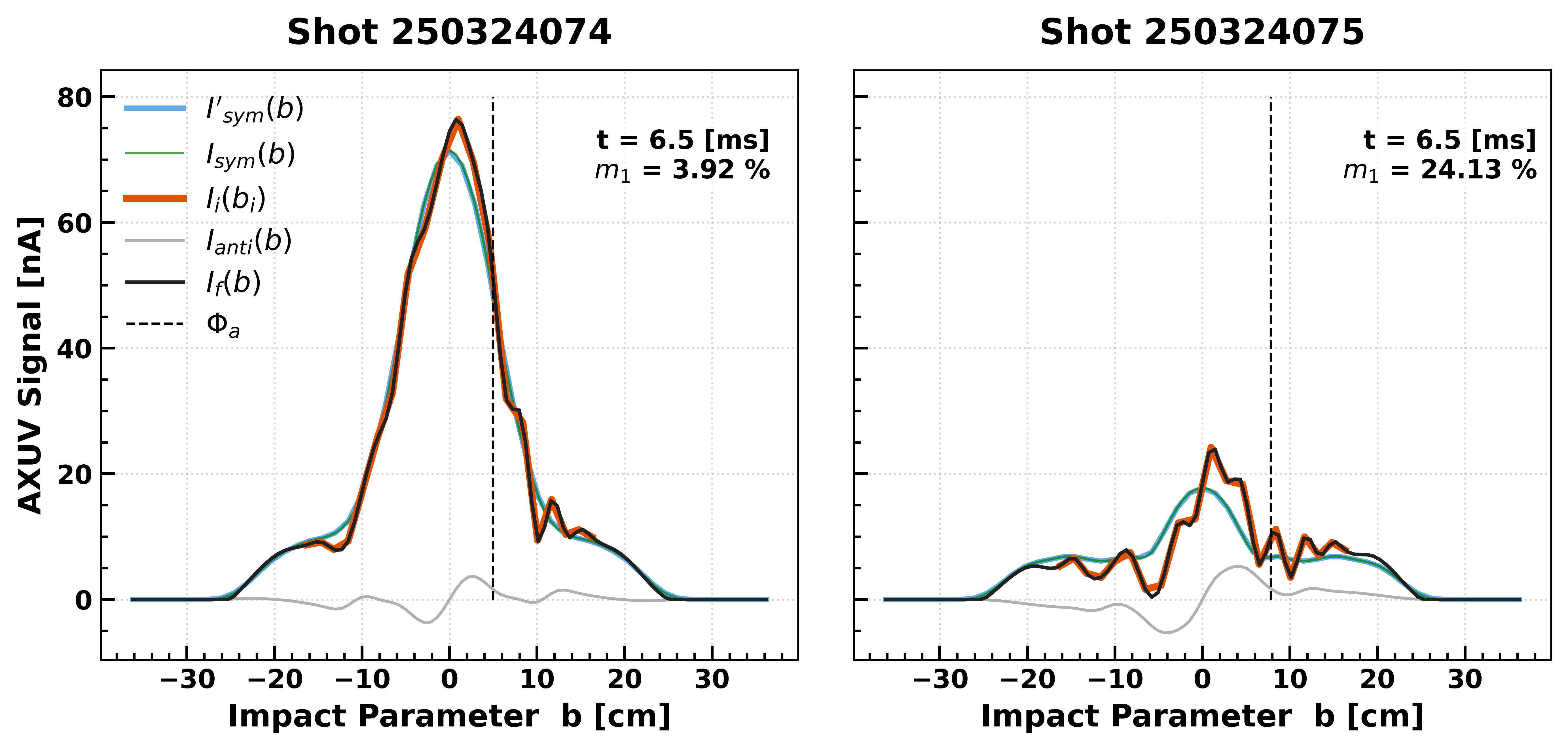}

\caption{\label{fig:abel_validation} Validation of the Abel inversion using representative WHAM discharges. The raw AXUV signals $I_{\mathrm{i}}(b_i)$ (orange), the reconstructed symmetrized component $I'_{\mathrm{sym}}(b)$ (blue), the measured symmetric signal $I_{\mathrm{sym}}(b)$ (green), and the anti–symmetrized signal $I_{\mathrm{anti}}(b)$ (gray) are shown for comparison. The black dashed line indicates the calculated anti–symmetry centroid $\Phi_{\mathrm{a}}$ for each discharge, together with the corresponding $m = 1$ population, illustrating its influence on plasma sustainment and emission asymmetry.}
\end{figure}

Figure~\ref{fig:abel_validation} illustrates the reconstruction and its validation for critical conduction discharges~74 and~75.  
Each panel shows the raw diode signals $I_i(b_i)$, their anti–symmetrized components $I_{\mathrm{anti}}(b)$, and the synthetic line–integrated signals $I'_{\mathrm{sym}}(b)$ obtained by forward–projecting the reconstructed $\bar{S}(\rho,t)$.  
The close agreement between $I_{\mathrm{sym}}(b)$ and the synthetic projections demonstrates that the Abel inversion captures the dominant symmetric emission.  
Deviations from perfect symmetry quantify the associated $m=1$ mode (via $I_{\mathrm{anti}}$ and $\Phi_{\mathrm{a}}$), clarifying its role in plasma sustainment.  
The measured–reconstructed consistency supports the use of $\bar{S}(\rho,t)$ for subsequent macroscopic analysis, including the moment–based quant
ities $\Phi(t)$, $R(t)$, and $\chi(t)$ introduced in Sec.~\ref{sec:stability}.

\section{\label{app:overview}AXUV Diagnostic Overview Plot}

\begin{figure*}[t!]
\centering
\includegraphics[width=\textwidth]{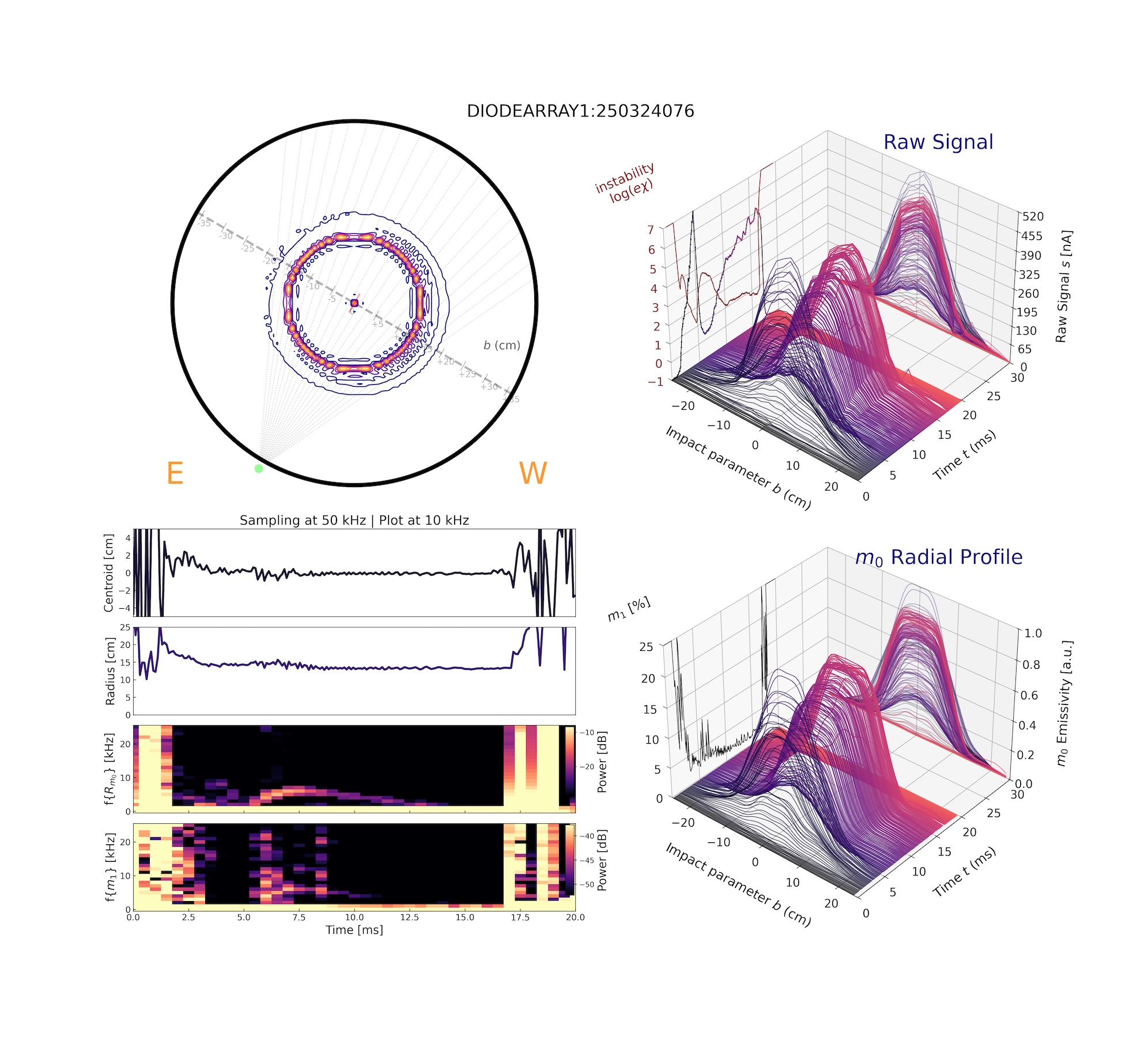}
\caption{\label{fig:axuv_overview}
Comprehensive overview of the single-array AXUV analysis for a representative WHAM discharge. Left column: (top) plasma centroid and radius trajectory relative to the WHAM vessel (black circle); (second) centroid position versus time; (third) effective radius versus time; (fourth) short-time Fourier transform (STFT) of the radius trace; (bottom) STFT of the inferred $m{=}1$ population amplitude. Right column: (top) raw diode-array signal evolution during the discharge, with total emissivity and $\chi(t)$ projected on the left wall and instantaneous signals on the front wall; (bottom) Abel-inverted emissivity evolution, with the $m{=}1$ population trace projected on the left wall and the instantaneous reconstructed emissivity on the front wall.The combined panels illustrate the linkage between measured radiation dynamics, macroscopic plasma motion, and instability evolution.}
\end{figure*}

Figure~\ref{fig:axuv_overview} presents a representative overview of the complete AXUV analysis workflow for a typical WHAM discharge.  
The calibrated 20-channel diode array signals are processed to yield both the time evolution of macroscopic emissivity moments and the derived instability parameter~$\chi(t)$.  
The figure combines geometric visualization, temporal traces, and spectral analyses to illustrate the information content extracted from a single AXUV array.

The left column summarizes the evolution of plasma geometry and macroscopic parameters.  
The top panel shows the plasma centroid and radius trajectory within the WHAM vessel cross-section (black circle) throughout the discharge.  
The next two panels display the corresponding time traces of the centroid position and effective radius, respectively.  
The lower panels present short-time Fourier transforms (STFTs) of the radius trace and of the inferred $m=1$ population amplitude, revealing the dominant fluctuation frequencies and their temporal evolution.

The right column provides complementary visualizations of the measured and reconstructed emissivity signals. The upper panel shows the raw 20-channel diode signals evolving in time, with the total emissivity and $\chi(t)$ projected onto the left wall and the instantaneous signals projected onto the front wall. The lower panel shows the Abel-inverted emissivity distribution evolving in time, with the corresponding $m=1$ population trace projected on the left wall and the instantaneous reconstructed emissivity projected on the front wall.  

Together, these visualizations demonstrate that the single-array AXUV diagnostic captures the full temporal and spatial evolution of plasma emissivity, including global motion, mode content, and macroscopic instability signatures, with sub-millisecond resolution.

\nocite{*}
\bibliography{AXUV_RSI}

\end{document}